\documentclass{Raju_et_al}
\usepackage{graphicx}

\usepackage{graphicx}
\usepackage{color}
\usepackage{hyperref}
\hypersetup{colorlinks=true,urlcolor=blue,citecolor=blue,linkcolor=blue,bookmarks=true,bookmarksopen=false,pagebackref=true}

\newcommand{\fig}[1]{Fig.~\ref{#1}}

\newcommand{\Neel}{N\'{e}el}
\newcommand{\nsk}{n_{\mathrm{sk}}} 
\newcommand{\Qsk}{Q_{\mathrm{sk}}} 
\newcommand{\nw}{n_{\mathrm{w}}} 
\newcommand{\nt}{n_{\mathrm{T}}} 

\newcommand{\qw}{Q_{\mathrm{W}}} 

\newcommand{\rhoth}{\rho_\mathrm{TH}} 
\newcommand{\Rhoth}{$\rhoth$}
\newcommand{\drho}{\Delta\rho_{yx}} 
\newcommand{\Drho}{$\drho$} 
\newcommand{\beff}{B_\mathrm{eff}}
\newcommand{\Beff}{$\beff$}

\newcommand{\Keff}{K_\mathrm{eff}}
\usepackage{xcolor}
\reversemarginpar
\title{Chiral magnetic textures in Ir/Fe/Co/Pt multilayers: Evolution and topological Hall signature}

\author{M. Raju$^{1\ast}$, A. Yagil$^{2\ast}$, Anjan Soumyanarayanan$^{3,1}$, Anthony K. C. Tan$^{3}$, A. Almoalem$^{2}$, O. M. Auslaender$^{2\dagger}$, \& C. Panagopoulos$^{1\dagger}$}

\begin{document}

\maketitle

\begin{affiliations}
 \item Division of Physics and Applied Physics, School of Physical and Mathematical Sciences, Nanyang Technological University, 637371 Singapore
 \item Department of Physics, Technion, Haifa 32000, Israel
 \item Data Storage Institute, Agency for Science, Technology and Research (A*STAR), 2 Fusionopolis Way, 138634 Singapore
\end{affiliations}

\begin{abstract}
\sloppy
Skyrmions are topologically protected, two-dimensional, localized hedgehogs and whorls of spin\cite{Nagaosa2013}. Originally invented as a concept in field theory for nuclear interactions\cite{Skyrme1961}, skyrmions are central to a wide range of phenomena in
condensed matter\cite{Sondhi1993,Senthil2004,Wang2015}. Their realization at room temperature (RT) in magnetic multilayers\cite{MoreauLuchaire2016,Woo2016,Soumyanarayanan2017} has generated considerable interest, fueled by technological prospects and the access granted to fundamental questions.   The interaction of skyrmions with charge carriers\cite{Bogdanov2001,Nagaosa2013,Soumyanarayanan2016a,Jiang2016,Litzius2017, Soumyanarayanan2017}  gives rise to exotic electrodynamics, such as the topological Hall effect (THE), the Hall response to an emergent magnetic field [$\mathbf{B_{eff}({r})}$], a manifestation of the skyrmion Berry-phase\cite{Neubauer2009,Lee2009}.   The proposal that THE can be used to detect skyrmions needs to be tested quantitatively. For that it is imperative to develop comprehensive understanding of skyrmions and other chiral textures, and their electrical fingerprint.   Here, using Hall transport and magnetic imaging, we track the evolution of magnetic textures and their THE signature in a technologically viable multilayer film as a function of temperature ($\mathbf{T}$) and out-of-plane applied magnetic field ($\mathbf{H}$).  We show that topological Hall resistivity ($\mathbf{\rho_{TH}}$) scales with the density of isolated skyrmions ($\mathbf{n_{sk}}$) over a wide range of $\mathbf{T}$, confirming the impact of the skyrmion Berry-phase on electronic transport. We find that at higher $\mathbf{n_{sk}}$ skyrmions cluster into worms which carry considerable topological charge, unlike topologically-trivial spin spirals. While we establish a qualitative agreement between $\mathbf{\rho_{TH}(H,T)}$ and areal density of topological charge  $\mathbf{n_{T}(H,T)}$, our detailed quantitative analysis shows a much larger $\mathbf{\rho_{TH}}$  than the prevailing theory predicts for observed $\mathbf{n_{T}}$.    Our results are pertinent for the understanding of skyrmion-THE  in multilayers, where interfacial interactions, multiband transport and non-adiabatic effects play an important role, and for skyrmion applications which rely on THE. The presence of worms composed of skyrmions highlights the role of skyrmion-skyrmion interactions, which are yet to be understood.  
\end{abstract}

  %
%

\sloppy
When charge carriers flow through a conductor with their spins tracking the skyrmion spin texture, the topological Hall resistivity, $\rhoth$, is predicted to be\cite{Nagaosa2013}:
\begin{equation}\label{eq:rhoth}
\rhoth=PR'_0\nt\Phi_0.
\end{equation}
Here $R'_0$ is usually taken to be the ordinary Hall coefficient $R_0$\cite{Neubauer2009,Matsuno2016,Maccariello2018,Zeissler2017}, $0<P<1$ is the spin-polarization of the charge carriers, $\nt$ the areal density of topological charge, and $\beff(\mathbf{r})$ is manifested through $\Phi_0=h/e$, the flux quantum ($h$ is Planck's constant, $-e$ is the electron charge). %
Assuming skyrmions are the sole carriers of topological charge $|\Qsk|=1$, $\nt$ is the density of skyrmions, $\nsk$. Thus, within the adiabatic approximation, one expects a straightforward  correlation between $\rhoth$ and $\nsk$. 
From the first observations in B20 systems\cite{Neubauer2009, Lee2009} to the recent multilayers,
THE has been used as an indicator for the presence of skyrmions\cite{Matsuno2016,Soumyanarayanan2017}.  %
However, a clear understanding of the effect is still lacking\cite{Denisov2017}, especially in technologically viable multilayer films\cite{Soumyanarayanan2017,Maccariello2018,Zeissler2017}, where disorder and interface effects can play an important role\cite{Woo2016,Yagil2017,Legrand2017b}.   

\sloppy
Using magnetic force microscopy (MFM) and transport measurements we present a comprehensive picture of the evolution of magnetic textures and their THE  signature in a multilayer film capable of hosting skyrmions from RT down to at least $5$~K. 
We demonstrate the relationship between $\nsk$ and $|\rhoth|$ over a $\approx200$~K temperature ($T$) range. %
As the applied field $H$ is swept towards zero, we find that skyrmions aggregate in worm-like magnetic textures, which carry a large topological charge, and manifest as peaks in \Rhoth. Quantitative modeling of these worm-textures uncovers qualitative agreement between $\rhoth(H,T)$ and $\nt(H,T)$. %
Despite this, we find a large quantitative discrepancy indicating that the effect in multilayers is more involved. %

\sloppy
Here we use sputtered $[$Ir(1)/Fe(0.5)/Co(0.5)/Pt(1)$]_\mathrm{20}$ (in  parenthesis  -- thickness in nanometers) multilayer films, with the composition chosen for exhibiting skyrmions across a large range of $T$. The RT characterization of the films through magnetization, MFM and micromagnetic simulations indicated the key 
magnetic parameters responsible for skyrmion formation, Dzyaloshinskii-Moriya interaction (DMI, $D$), and exchange interaction ($A$) to be $D\approx2.0~\mathrm{mJ/m^2}$, $A\approx11$~pJ/m\cite{Soumyanarayanan2017}. The effective anisotropy ($\Keff$)  varies in the range $\approx0.2$--$0.01~\mathrm{MJ/m^3}$ as we change $T$ from $5$~K to $300$~K\cite{SuM}.   As demonstrated here, control over skyrmion density through variation of $T$ is the key for unambiguous verification of the skyrmion THE signature.   %
In contrast to the  B20 compounds, which host lattices of  tubular Bloch-skyrmions\cite{Milde2013}, multilayers sustain skyrmions with tunable diverse properties, and offer smoother integration  with existing spintronic technologies. Spin textures in multilayers are influenced by interlayer dipolar and exchange interactions, magnetic frustration\cite{Rozsa2017}, and granularity\cite{Woo2016}, which can pin, stabilize, and deform the spin textures, and result in   coupled pancake-skyrmions with different topologies\cite{Rozsa2017,Legrand2017b}. This complexity, and associated tunability, provide means for exploring the interplay between disorder, interactions, and topology.

\begin{figure*}
	\centering
	\includegraphics[width=0.95\linewidth]{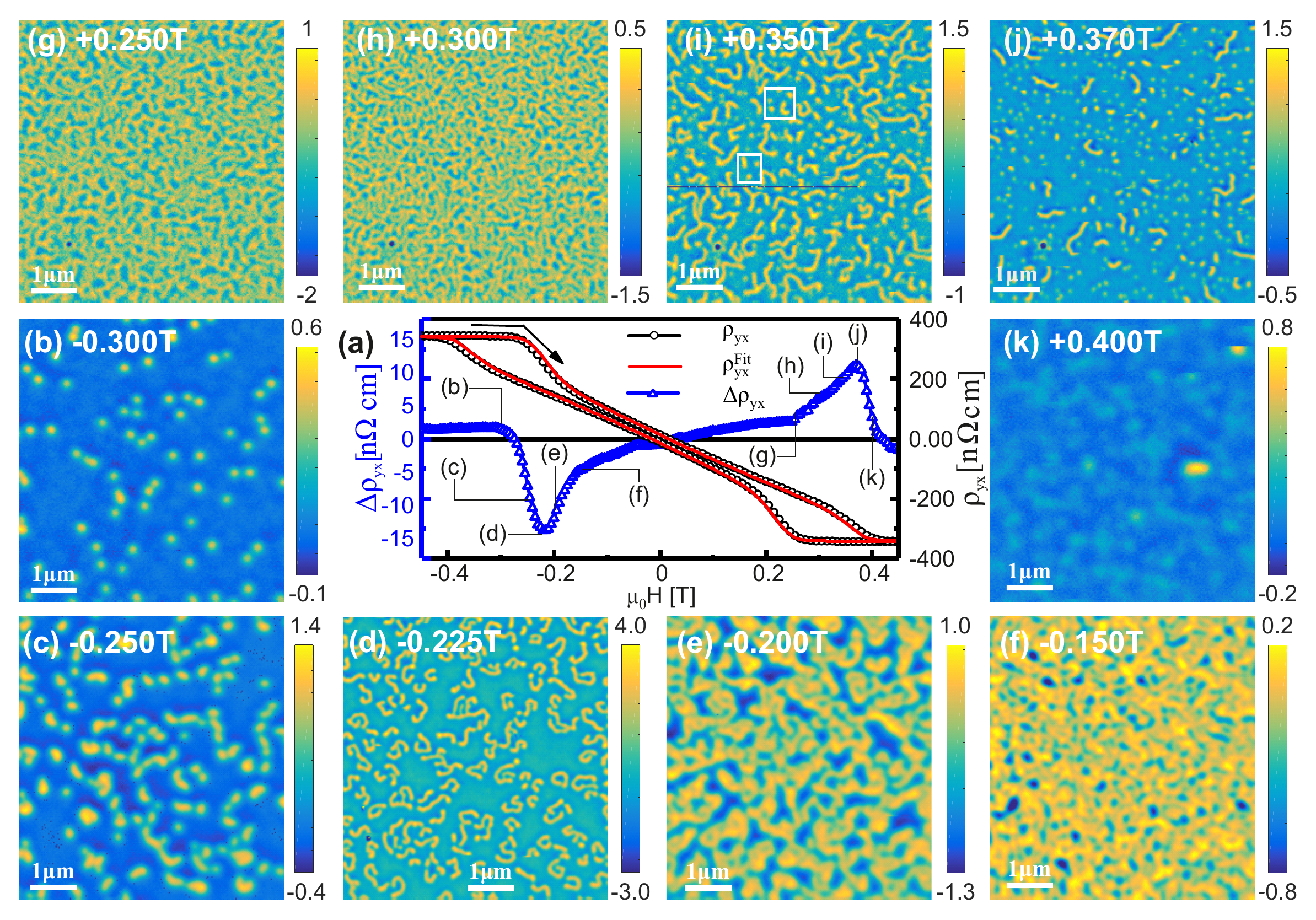}
	\caption[Comparison of the field evolution of MFM images with THE at $T=5$~K]{Evolution of magnetic textures and THE with $H$ at $T=5$~K.  %
		\textbf{(a)} $\rho_{yx}(H)$, $\rho^{fit}_{yx}(H)$ and the residual $\drho(H)=\rho_{yx}(H)-\rho^{fit}_{yx}(H)\approx\rhoth(H)$. Black arrow indicates field sweep direction for \Drho\ and MFM, while $\rho_{yx}$ and $\rho^{fit}_{yx}$ are shown for both sweep directions.  %
		\textbf{(b)-(k)} Selected MFM scans [full sequence in\cite{SuM}, scan-height $h = 75, 60, 40, 60, 65, 50, 60, 50, 45, 60$~nm for each of (b)-(k); color bars give $\Delta f$ range.]  %
		Frames in (i) mark features in focus in \fig{fig:worm_and_dense}(b)-(g).
	}
	\label{fig:MFM_5K}
\end{figure*}

\sloppy
The magnetoresistance and Hall effect were measured using a lock-in with non-perturbative current densities ($\approx10^5~\mathrm{A/m^2}$).  The presence of skyrmions is associated with  an additional component in the measured Hall signal ($\rho_{yx}$)\cite{Neubauer2009,Lee2009,Matsuno2016}. This contribution can be quantified by resolving $\rho_{yx}$  into the ordinary ($R_0H$) and anomalous [$R_\mathrm{S}M(H)$] Hall components, and an extra component (\Rhoth)\cite{Neubauer2009,Matsuno2016,Maccariello2018,Zeissler2017}:
\begin{equation}\label{eq:AHE}
\rho_{yx}(H)= R_0H+R_\mathrm{S}M(H)+\rhoth(H).
\end{equation}
We estimate $\rhoth(H)$ by $\drho(H)$, the residual of the fit of $\rho_{yx}(H)$ to $\rho_{yx}^{fit}(H)=R_0H+R_\mathrm{S}M(H)$, which also yields $R_0$ and $R_\mathrm{S}$\cite{SuM}. The accuracy of \Drho\ is ensured by calibrating field offsets to avoid artifacts resulting from using different measurement set-ups\cite{SuM}. %
Our conservative estimate for the overall error in \Drho, including a contribution from data analysis, is $\pm2~\mathrm{n\Omega\cdot cm}$, corresponding to the non-zero residual signal beyond saturation, where there are no skyrmions. %
Figure~\ref{fig:MFM_5K}(a) shows $\drho(H)$ at $5$~K, with the overall features persisting to at least $300$~K\cite{Soumyanarayanan2017}.%

\sloppy
For MFM we followed the same field sweep from  saturation as we did for $\rho_{yx}(H)$ and $M(H)$ -- the images were acquired at field increments as $H$ was swept\cite{SuM}. Figure~\ref{fig:MFM_5K}(b)-(k) shows the result for $5$~K. Overall, we observe a similar evolution of the magnetic textures at $T=50,~100,~150,~200$~K\cite{SuM}.

\sloppy
We begin by comparing the  magnetic textures to $\drho(H)$. Beyond saturation MFM shows the null signal expected for a polarized ferromagnet\cite{SuM}. This corresponds to suppressed \Drho\ [\fig{fig:MFM_5K}(a)], indicating the topologically trivial nature of the polarized state. The onset of \Drho\ commences at $\mu_0H\approx-0.3$~T with the nucleation of sub-$100$~nm  magnetic domains, indicating the emergence of a finite Berry-phase resulting from nontrivial topology [Fig.~\ref{fig:MFM_5K}(b)], which we  identify as \Neel-skyrmions\cite{Yagil2017}.  %
By $\mu_0H\approx-0.25$~T [\fig{fig:MFM_5K}(c)] the increasing $\nsk$ corresponds to a substantial $\drho$ as expected from $|\rhoth|\propto\nsk$\cite {Nagaosa2013}. Also, as $\nsk$ increases, skyrmions aggregate to form worm-like features [\fig{fig:MFM_5K}(c)]. By $\mu_0H\approx-0.225$~T, images show only worms [\fig{fig:MFM_5K}(d),(e)]. Surprisingly, at this field \Drho\ peaks, indicating a significant contribution from the worms which therefore must have nontrivial topology. %
Meanwhile, the dense textures at intermediate $H$, [Fig.~\ref{fig:MFM_5K}(f)-(h),\cite{SuM}], correspond to reduced, yet finite, \Drho. %
Careful inspection of such scans also reveals worm-like features\cite{SuM}, to which we attribute the finite magnitude of \Drho.

\sloppy
As  $H=0$ is approached, the worms  evolve into labyrinthine helical stripes, and proliferate at the expense of the polarized background. This is coincident with a suppression of \Drho, highlighting the close relationship between \Drho\ and the magnetic texture. As $H$ is increased towards positive saturation, the labyrinthine stripes evolve into worms, skyrmions, and eventually a uniformly polarized phase [Fig.~\ref{fig:MFM_5K}(i)-(k)]. As expected for a texture with an opposite topological charge, the sign of $\drho$ is reversed when $H>0$. However, the MFM contrast does not change, due to reversal of the tip magnetization near $0.1$~T. %

\sloppy
The distinct field ranges for isolated skyrmions (near saturation), worms (negative peak in \Drho) and their coexistence (positive peak in \Drho) offer a unique opportunity to compare the magnetic texture with \Drho. In particular: (i) Does \Drho\ track $\nsk$? (ii) How do worms produce such a large \Drho? (iii) Is there quantitative consistency between \Drho\ and $\nt$?   

\begin{figure*}
	\centering
	\includegraphics[width=1\linewidth]{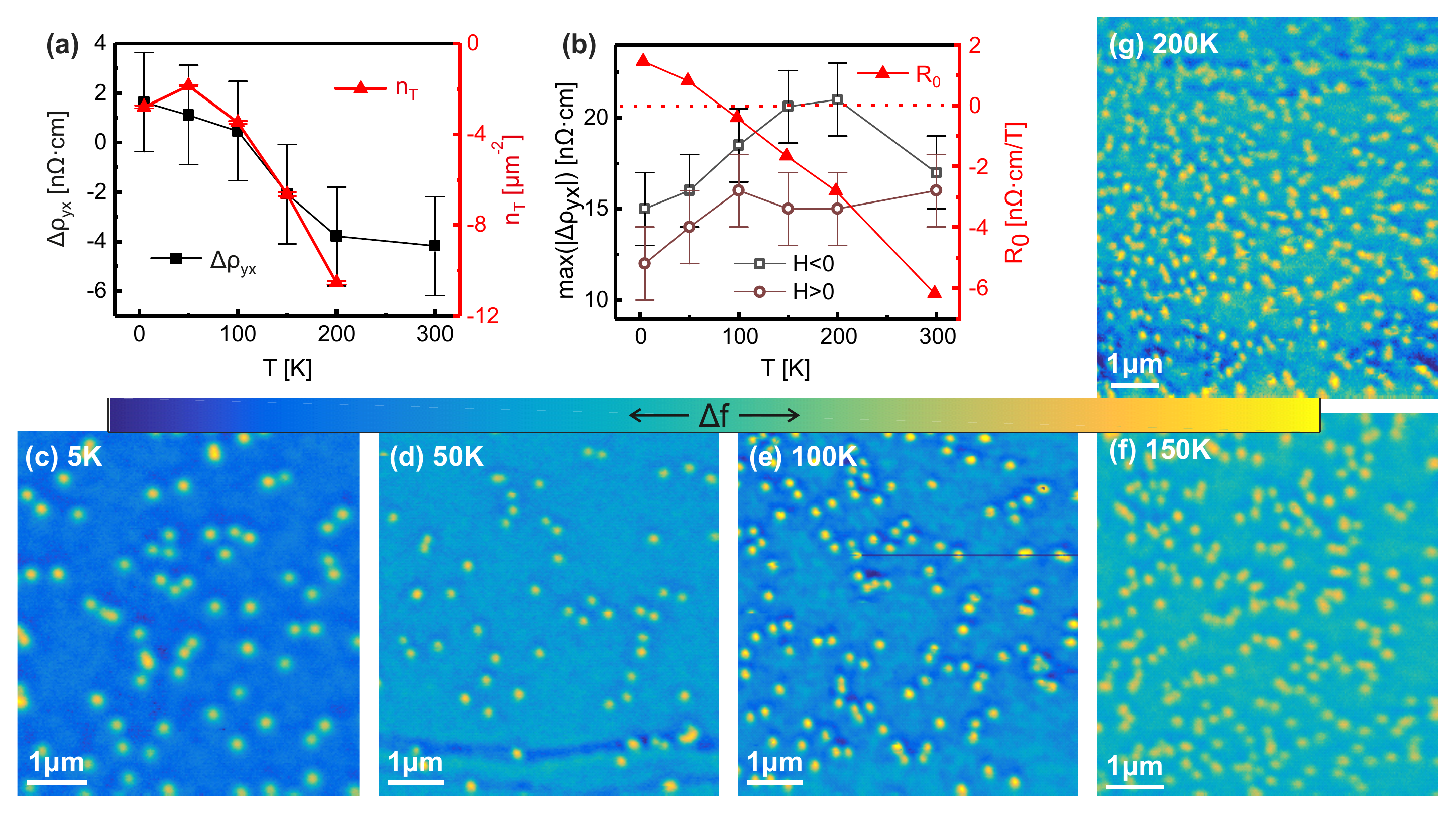}
	\caption[Temperature dependence of $\nt$, $\drho$, and $R_0$.]{%
		Temperature dependence of $\nt$, $\drho$, and $R_0$. %
		\textbf{(a)} $\drho(T)$ (left axis, squares) compared with topological charge $|\nt(T)|=\nsk(T)$ (right axis, triangles) at $-0.3$~T, corresponding to the isolated skyrmions in (c)-(g). The error bars for $\drho$ represent a conservative estimate of the systematic error. 	 %
		\textbf{(b)} Left axis: Magnitude of the THE peaks ($H<0$ -- squares, $H>0$ -- circles). Right axis: The fit parameter $R_0(T)$ from Eq.~\ref{eq:AHE} (triangles). %
		\textbf{(c)-(g)} MFM scans showing isolated skyrmions for $\nt(T)$ in (a). %
		[For (c)-(g) scan-height $h=75,40,40,100,40$~nm, $\Delta f_\mathrm{range}=0.7,6,4,1.8,2$~Hz. (c) is Fig.~\ref{fig:MFM_5K}(b).]
	}
	\label{fig:Tdep_fig}
\end{figure*}

\sloppy
To address (i) we exploit $\nsk(T)$, which increases by an order-of-magnitude when we increase $T$ from $5$~K to $200$~K (Fig.~\ref{fig:Tdep_fig}). We attribute this proliferation  to the suppression of the critical DMI $D_\mathrm{C}=4\left(A\Keff\right)^{1/2}/\pi$ as $\Keff$ decreases\cite{Soumyanarayanan2017}. Interestingly,  skyrmion size is only weakly dependent on $T$, not unlike theoretical predictions\cite{Tomasello2017}. Importantly, we find that $\drho(T)$ tracks $\nsk(T)$ over the entire range [Fig.~\ref{fig:Tdep_fig}(a)]. This verification of $|\rhoth|\propto\nsk$ was not possible previously due to limited tunability of $\nsk$ in skyrmion systems reported so far. This validates the topologically non-trivial nature of the skyrmions and the viability of our approach.    

\begin{figure*}
	\centering
	\includegraphics[height=0.33\linewidth]{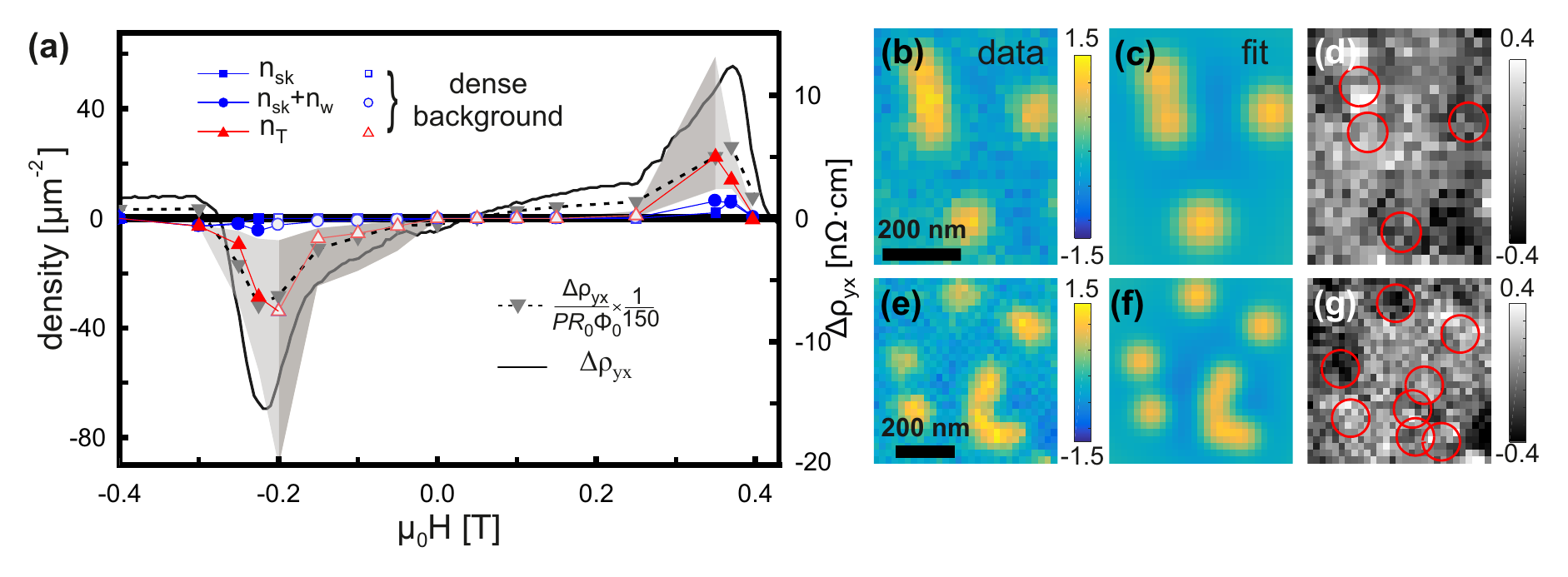}
	\caption[Comparison between \Drho\ and the signed density of topological charge.]{Comparison between \Drho\ and the signed density of topological charge at $5$~K. %
		\textbf{(a)} Right axis: \Drho\ (solid line). Left axis: $\nsk$ (squares), $\nsk+\nw$ (circles), $\nt$ (triangles), and $\drho/(PR_0\Phi_0)$ (cf. Eq.~\ref{eq:rhoth}, inverted  triangles with dotted line) using $P=0.56$\cite{Rajanikanth2010} and $R_0$ from Fig.~\ref{fig:Tdep_fig}(b). For $\nt$, each worm is assigned several skyrmions $|\qw|$ by fit [cf. (b)-(g)], and each isolated skyrmion is counted once.  %
		Empty symbols indicate points with a lower confidence, that result from counting worms in a dense background\cite{SuM}. %
		Shaded area shows confidence bounds for $\nt$ resulting from fit details discussed in\cite{SuM}.  %
		\textbf{(b),(e)} Zooms  on areas in Fig.~\ref{fig:MFM_5K}(i) with skyrmions and isolated worms. %
		\textbf{(c),(f)} Results of MEFS fits wherein all skyrmions, including those assigned to worms, are treated as identical. %
		The fit peak height $\approx1.15$~Hz, full width at half maximum (FWHM) $\approx100$~nm, and $|\qw|=2$ in (c) and $|\qw|=4$ in (f).] 
		\textbf{(d),(g)} Difference plots between (b) and (c), and between (e) and (f), showing the quality of the fit. Circles give the locations of the skyrmions in the fit. [Color bars give $\Delta f_\mathrm{range}$.] 		%
	}
	\label{fig:worm_and_dense}
\end{figure*}

\sloppy
Having established the direct correspondence between $\nsk$ and $\drho$, we now examine the worms. Though the presence of worms is expected in systems with competing interactions, such as in ferromagnetic films\cite{Seul1995}, their topological role is not obvious.  %
In recent MFM work on Bloch skyrmions in a B20 compound\cite{Milde2013}, helical stripe domains resulting from merging skyrmions were assigned $|\qw|=1$ -- each stripe was described by two half-skyrmions connected by a topologically trivial straight domain\cite{Milde2013}.
This motivates the examination of whether worms carrying a topological charge $|\qw|=1$ can describe our results.
We therefore plot $\nsk+\nw$ ($\nw$ is the number of worms per unit area) in Fig.~\ref{fig:worm_and_dense}(a), with the sign chosen from the sign of \Drho. %
As the plot shows, both $\nsk(H)$ and $\nsk(H)+\nw(H)$ do not track $\drho(H)$. This calls for a closer look at the topological nature of the worms.

\sloppy
The following analysis is motivated by sequences like \fig{fig:MFM_5K}(b)-(d), which suggest that worms result from skyrmions clustering as $\nsk$ increases. The transition of worms into typical stripe domains with $|\qw|=1$ requires a complete unwinding of their internal spin structure. The energy barrier for this suggests that the effective topological charge should be at least equal to the total number of skyrmions that form a worm (i.e. $|\qw|>1$).  While skyrmions are expected to repel each other on very short length-scales because of exchange coupling\cite{Lin2013}, clusters can be stabilized by attraction on an intermediate scale, due to exchange frustration \cite{Rozsa2016}.

\sloppy
Here our recent work, a magnetic multipole expansion of the field from skyrmions (MEFS)\cite{Yagil2017}, provides a direct method to associate an effective $\qw$ with each worm.      %
Figures~\ref{fig:worm_and_dense}(b)-(g) show two typical examples where we fit the measured signal from worms by trains of skyrmions. %
Such images, which contain both skyrmions and worms, provide the foundation for this kind of analysis -- the isolated skyrmions serve to constrain the fit amplitude per skyrmion, and improve the accuracy\cite{SuM}. %
Meanwhile, the analysis of images containing only worms [Fig.~\ref{fig:MFM_5K}(d)] hinges on skyrmions-skyrmion repulsion on a length scale comparable to their radius\cite{Lin2013}. Therefore, the number of skyrmions clustered in a worm is determined by the total length of the worm, and the typical radius of skyrmions [$\approx40$~nm\cite{Yagil2017}]. For images with densely packed features [e.g. Figs.~\ref{fig:MFM_5K}(f)-(h)]  identifying and extracting the worms themselves requires additional image processing, for which we employ a deep-learning-model-based algorithm, which extracts features relevant for classification from supplied examples\cite{SuM}.

\sloppy
The qualitative match between $\nt(H)$ estimated from $\sum\qw+\sum\Qsk$ and $\drho(H)$  [\fig{fig:worm_and_dense}(a)] reinforces our modeling of worms as trains of skyrmions and confirms that $\drho\approx\rhoth$, and with it the topological nature of the worms. Our observations on the emergence of worms with a high topological number, previously not noticed experimentally, indicate they are distinct from trivial spin spirals, and form an essential part of the phase diagram for multilayer skyrmions\cite{Dupe2014}. The presence of worms in a tunable multilayer offers a platform for studying skyrmion-skyrmion interactions over a wide parameter range, as well as applications, such as skyrmion racetracks\cite{Woo2016}.

\sloppy
Having established the nontrivial topology of the worms and a qualitative match between $\rhoth(H,T)$ and $\nt(H,T)$,  we examine the quantitative match. As we show in Fig.~\ref{fig:worm_and_dense}(a), the density of topological charge estimated from THE [$\drho/(PR_0\Phi_0)$, Eq.~\ref{eq:rhoth}] indicates a two-orders-of-magnitude discrepancy.   
This implies that the assumptions invoked in using Eq.~\ref{eq:rhoth} to understand the topological signatures of chiral magnetic textures in multilayer skyrmion-hosts, are insufficient for a comprehensive description of the measured \Drho. 

\sloppy
A possible culprit is the assignment $R'_0=R_0$  in Eq.~\ref{eq:rhoth}\cite{Onoda2004}, justified for a single band material. This is not the case here: Bulk Fe and Co, the ferromagnetic ingredients of our multilayer stack, have several active electron and hole bands\cite{Pugh1953}.   %
In such materials $R_0$ is suppressed because electrons and holes, which experience the same $H$, compensate each other's contributions. The cancellation estimated from values reported for bulk Fe\cite{Pugh1953} indicate suppression of $R_0$ by an order of magnitude from the separate contributions of individual bands, partially addressing the discrepancy\cite{SuM}. Importantly, the cancellation may not happen in the same way for \Beff\ -- the Berry-phase can act differently on charge carriers from different bands\cite{Ritz2013a}, with an associated sensitivity to occupation\cite{Gobel2017}. %
The fact that the peak value of $\drho(H)$ changes by only $\approx25\%$ with $T$ despite the sign change of $R_0(T)$ [\fig{fig:Tdep_fig}(b)], probably because of small variations of the occupations of the compensating bands, further confirms that $R_0'\neq R_0$. In this case, using $R_0$ in Eq.~\ref{eq:rhoth} underestimates $\rhoth$\cite{Maccariello2018,Zeissler2017}, although once $R_0'$ is determined this fundamental equation may still be used. For this it is essential to account for the electronic band structure of the chiral magnet.

\sloppy
Our complementary imaging and electrical transport studies elucidate the complexity of the Berry-phase associated with the electrical fingerprint of chiral magnetic textures in technologically viable magnetic multilayers. For a comprehensive understanding and in order to utilize THE emerging from magnetic skyrmions, it is imperative to consider (a) the band structure contributing to THE, (b) the possibility of $|\Qsk|>1$\cite{Rozsa2017}, (c) the coupling of skyrmions across layers and complex magnetic textures in buried interfaces\cite{Legrand2017b}, (d) the contribution from topologically trivial chiral configurations driven by magnetic spin-frustration\cite{Denisov2017}, and (e) the validity of the commonly assumed adiabatic approximation\cite{Denisov2017}.
\begin{addendum}
	\item It is our pleasure to thank A. Petrovi\'{c} for inputs on transport experiments, as well as D. Arovas, A. Auerbach, Shi-Zeng Lin, D. Podolsky, M. Reznikov, and A. Turner for illuminating discussions. We also thank G. Goldman and  Y. Schechner for help with image analysis. %
	The work in Singapore was supported by the Ministry of Education (MoE) -- Academic Research Fund (Ref. No. MOE2014-T2-1-050), the National Research Foundation	-- NRF Investigatorship (Reference No. NRF-NRFI2015-04), and the A{*}STAR Pharos Fund (1527400026). 		Work at Technion was supported by the Israel Science Foundation (Grant no. 1897/14). We would also like to	thank the Micro Nano Fabrication Unit, and to acknowledge support from the Russell Berrie Nano Technology Institute, both at the Technion.
	\item[Contributions] MR, AS, OMA, and CP designed and initiated the research.
	MR deposited the films and characterized them with AS and AT. MR performed and analyzed magnetization and transport measurements with inputs from AS, OMA, and CP. AY performed the low temperature MFM experiments with assistance from AA, and AY and OMA analyzed the low temperature MFM data. OMA and CP coordinated the project. All authors discussed the results and provided inputs to the manuscript.
	\item[$^\ast$These authors are equal contributors] 
	\item[Competing Interests] The authors declare that they have no competing financial interests.
	\item[$^\dagger$Correspondence] Correspondence and requests for materials
	should be addressed to to: ophir@physics.technion.ac.il / christos@ntu.edu.sg.
	
\end{addendum}

\section*{References}




\newpage

\end{document}